\documentclass{article}
\usepackage{spconf,amsmath,epsfig,amsfonts,subfigure,graphicx,amssymb}
\usepackage{enumitem}

\title{SETI DETECTION STRATEGIES FOR SINGLE DISH RADIO TELESCOPES}
\name{Gregory Hellbourg}
\address{Berkeley SETI Research Center - University of California - Berkeley, CA, USA}

\begin{document}

\maketitle

\begin{abstract}
Radio Searches for Extra Terrestrial Intelligence aim at detecting artificial transmissions from extra terrestrial communicative civilizations. The lack of prior knowledge concerning these potential transmissions increase the search parameter space. Ground-based single dish radio telescopes offer high sensitivity, but standard data products are limited to power spectral density estimates.

To overcome important classical energy detector limitations, two detection strategies based on asynchronous \emph{ON} and \emph{OFF} astronomical target observations are proposed. Statistical models are described to enable threshold selection and detection performance assessment.
\end{abstract}


\section{Introduction}
\label{sec:intro}

The Search for Extra Terrestrial Intelligence (SETI) at radio frequencies aims at finding the evidence of intelligent and communicative extra-terrestrial civilizations through the detection and localization of artificial\footnote{i.e. non-naturally produced signals, as suggested by our current understanding of astrophysics.} electromagnetic transmissions \cite{tarter2001search}. The expected signal might either correspond to a dedicated signaling beacon, or an
information-bearing radio transmission leakage. The absence of prior knowledge concerning extra terrestrial transmissions (including their existence in a first place) necessitates the exploration of a wide range of frequencies, epochs, directions-of-arrival, and signal characteristics to explore.

Ground-based searches allow higher amounts of data collection and processing than space-based searches. An important disadvantage of ground-based SETI is the presence of man-made Radio Frequency Interference (RFI) potentially mimicking the expected signal-of-interest.
While array telescopes offer a better tolerance to RFI \cite{hellbourg2014radio,perley2002attenuation}, single dish instruments remain easier to calibrate and manipulate. Their intrinsic directionality prevent spatially-blind surveys but enable high sensitivity targeted surveys such as \cite{turnbull2003target,isaacson2017breakthrough}.

This paper provides a statistical SETI data model for single dish telescope experiments, and compares two signal detection strategies.
Section \ref{sec:datamodel} describes the single dish SETI data model. Section \ref{sec:detect} addresses the SETI detection problem. Section \ref{sec:detectors} describes two detection strategies based on standard astronomical data products. Finally, section \ref{sec:conclusion} concludes this paper.

\section{Data model}
\label{sec:datamodel}

A single receiver radio telescope signal $x(t)$ at a given frequency during a SETI experiment is potentially composed of three main contributors : the extra terrestrial signal-of-interest $x_{\text{ET}}(t)$, the man-made interference $x_{\text{RFI}}(t)$, and the system noise $x_{\text{noise}}(t)$. This section models each of these contributors.

\subsection{System noise}

The system noise is the sum of various naturally-occurring independent centered stochastic processes
\cite{wilson2009tools}. As a consequence of the Central Limit Theorem, the system noise is modeled as a centered and complex normally distributed random variable $x_{\text{noise}}(t)\sim\mathcal{NC}(0,\sigma_n^2)$. We further assume the noise to be temporally and spatially white and stationary over the observation duration (of the order of minutes).

\subsection{Radio Frequency Interference}

Man-made RFI usually follow deterministic models (information bearing modulated signals). Observed over a narrow frequency bandwidth however, two cases are considered:

\subsubsection{Wide band RFI}
\label{subsub:widebandrfi}

When the RFI frequency bandwidth $\Delta f_{\text{RFI}}$ is much larger than the analysis frequency bandwidth $\delta f$ (i.e. $\delta f \ll \Delta f_{\text{RFI}}$), its contribution is modeled as a centered, stationary and complex normally distributed random variable $x_{\text{RFI}_{\text{wide}}}(t) \sim \mathcal{NC}(0,\sigma_r^2)$ \cite{rosenblatt1961some}, where $\sigma_r^2$ is the RFI power. $x_{\text{RFI}_{\text{wide}}}(t)$ takes in account the contribution of all RFI emitters at a given time and frequency locus.

\subsubsection{Narrow band RFI}

If the RFI bandwidth is matching or narrower than the analysis bandwidth, its contribution $x_{\text{RFI}_{\text{narrow}}}(t)$ is modeled as a square integrable function with energy $E_{\text{RFI}}$ over the observation duration.

\subsection{ET transmission}

Similarly to RFI, an extra terrestrial transmission can either be information-bearing and spread in frequency, or a simple narrow band beacon transmission.

\subsubsection{Wide band model}

A wide band information-bearing extra terrestrial transmission ($\delta f \ll \Delta f_{\text{ET}}$ with $\Delta f_{\text{ET}}$ the transmission bandwidth) is modeled as a centered, stationary and complex normally distributed random variable : $x_{\text{ET}_{\text{wide}}}(t)\sim\mathcal{NC}(0,\sigma_{\text{ET}}^2)$ (see section \ref{subsub:widebandrfi}).

\subsubsection{Narrow band model}

\begin{figure}
\centering
\includegraphics[width=.4\textwidth]{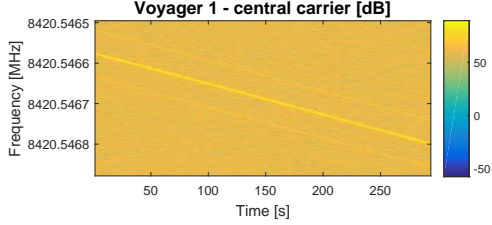}
\caption{Spectrogram of the central carrier of the Voyager 1 spacecraft telemetry transmission observed with the Green Bank Telescope (WV, USA) on September 19, 2016.}
\label{fig:voyager}
\end{figure}

SETI commonly searches for continuous wave extra terrestrial transmissions as no natural process seem to produce such signal, and they remain unaffected by interstellar/galactic scattering and scintillation \cite{ekers2002seti}. The relative motion between the extra terrestrial transmitter and the telescope (dominated by the Earth rotation and possibly the emitter intrinsic speed) induces however a Doppler frequency drift. The transmission is therefore modeled as a chirp. Figure \ref{fig:voyager} shows the central continuous wave carrier of the telemetry transmission from the Voyager 1 spacecraft \cite{ludwig2002j} on September 19, 2016, and illustrates the effect of the Earth rotation on the received signal.


\section{Extra Terrestrial Transmission Detection}
\label{sec:detect}

\subsection{Problem formulation}

At a given frequency, the SETI detection problem is formulated as a binary hypotheses testing problem with the following two hypotheses\footnote{$x_{\text{RFI}}(t) = 0$ in an RFI-free scenario.}:

\begin{equation*}
x(t)  = \begin{cases}
x_{\text{RFI}}(t) + x_{\text{noise}}(t) & \emph{H0}\\
x_{\text{ET}}(t) + x_{\text{RFI}}(t) + x_{\text{noise}}(t) & \emph{H1}
\end{cases}
\end{equation*}
where \emph{H0} corresponds to the absence of extra terrestrial transmission, while \emph{H1} stands for the complementary hypothesis.

The performance of a detector enabling the decision-taking process given an astronomical observation $x(t)$ is quantified
according to the probability of detection (or sensitivity) $\mathcal{P}_d = \mathcal{P}\left(\emph{H1}|\emph{H1}\right)$, and probability of false alarm $\mathcal{P}_{fa} = \mathcal{P}\left(\emph{H1}|\emph{H0}\right)$ \cite{van2004detection}. An efficient binary detector maximizes $\mathcal{P}_d$ while minimizing $\mathcal{P}_{fa}$ for a given data model.

\subsection{Single receiver telescope data product}
\label{sub:dataproduct}

The analog signal collected by a single dish radio telescope goes through a standard signal processing chain involving amplification, filtering, basebanding, digitization and channelization \cite{wilson2009tools}. Although artificial signal detection is better achieved when exploiting signal construction features such as their phase information or cyclostationarity \cite{4796930,4221497,maccone2010klt}, standard radio telescope only provide energy information as their main purpose is the recovery of natural stochastic sources.

At a given frequency resolution ($\delta f \approx$ 1 Hz), the standard astronomical data product
consists in an estimate of the received signal power over $N$ samples:

\begin{equation}
\hat{\sigma}_{xx^{*}}^2 = \frac{1}{N} \sum_{k=0}^{N-1} x[k \cdot T_s] x^{*}[k \cdot T_s]
\label{eq:estimate}
\end{equation}
where $T_s = \frac{1}{\delta f}$ is the sampling period and $(.)^{*}$ stands for the complex conjugate operator. We denote $n = k \cdot T_s$

The independence between $x_{\text{ET}}[n]$, $x_{\text{RFI}}[n]$, and $x_{\text{noise}}[n]$ leads to the following formulations of the \emph{H1} hypothesis according to section \ref{sec:datamodel} (\emph{H0} is deduced from these formulations, accounting for the independence between signals):

\begin{itemize}[noitemsep]
\item Wide band RFI / Wide band ET:\\
$x[n]\sim \mathcal{NC}(0,\sigma_{\text{ET}}^2+\sigma_r^2+\sigma_n^2)$
\item Wide band RFI / Narrow band ET:\\
$x[n]\sim \mathcal{NC}(x_{\text{ET}_{\text{narrow}}}[n],\sigma_r^2+\sigma_n^2)$
\item Narrow band RFI / Wide band ET:\\
$x[n]\sim \mathcal{NC}(x_{\text{RFI}_{\text{narrow}}}[n],\sigma_{\text{ET}}^2+\sigma_n^2)$
\item Narrow band RFI / Narrow band ET:\\
$x[n]\sim \mathcal{NC}(x_{\text{ET}_{\text{narrow}}}[n] + x_{\text{RFI}_{\text{narrow}}}[n],\sigma_n^2)$
\end{itemize}

The resulting distributions of the power estimates (equation \ref{eq:estimate}) of these data models follow standard
central and non-central $\chi^2$ distributions with $N$ degrees of freedom derived from the above distributions (see section \ref{sec:onoff}).

\subsection{Observations on Energy Detection}
\label{sub:observations_energy}

The classical energy detection is optimum in the absence of prior knowledge concerning the signal of interest $x_{\text{ET}}[n]$ \cite{van2004detection}. Its implementation requires
however an accurate knowledge of the \emph{H0} hypothesis distribution, i.e. system noise and RFI properties for the given data model. A mismatch between observed and modeled data \emph{H0} leads to the \emph{SNR wall} effect \cite{mariani2011snr,tandra2008snr} and reduced detection performances. The energy detector is therefore inadequate for radio astronomical applications for the following reasons:

\begin{itemize}[noitemsep]
\item The RFI environment is variable in time and frequency, and unpredictable. Interferers appear at various Interference-to-Noise Ratios (INR) depending on their original transmission power and the angle at which they impinge the telescope \cite{hellbourg2014radio,fridman2001rfi}.
\item A tracking dish telescope is in continuous motion to compensate for the Earth rotation. Its elevation therefore continuously varies, leading to an additional varying noise term originating from radio reflexions from the ground called \emph{spillover} noise \cite{srikanth1991comparison}.
\item Varying uncalibration : various factors affect the calibration of the instrument, such as the ambient temperature, power outages, or electronic stability \cite{masters2011calibration}.
\end{itemize}

Alternatives to the energy detection in unknown \emph{H0} environment include the ``\emph{ON}-\emph{OFF}'' and the $\mathcal{F}$-ratio tests.

\section{Energy-based detectors for single dish SETI}
\label{sec:detectors}

A single dish radio telescope is intrinsically limited to steer at one direction at any time. Such instrument is therefore not adapted to perform instantaneous blind all-sky surveys. However, their high directivity and sensitivity makes it an appropriate instrument for conducting targeted surveys (e.g. focusing on stars potentially hosting suitable bodies for the development of intelligent life in their neighborhood \cite{isaacson2017breakthrough,ekers2002seti}).

\subsection{Detector construction}

To overcome the limitations of the energy detector
described in section \ref{sub:observations_energy}, the proposed detectors are constructed with two statistically independent observations\footnote{$x_{\text{ON}}[n]$ and $x_{\text{OFF}}[n]$ are asynchronous observations. The two data sets are statistically independent due to the spatial and temporal independence and whiteness of the system noise, and the temporal independence of wide band sources.}:

\begin{itemize}[noitemsep]
\item An \emph{ON}-target measurement $x_{\text{ON}}[n]$, steering at a selected target. $x_{\text{ON}}[n]$ follows either the \emph{H0} or \emph{H1} hypothesis described in section \ref{sub:dataproduct}.
\item An \emph{OFF}-target measurement $x_{\text{OFF}}[n]$, steering away from the original target. The angular offset from the target must be small enough to ensure similar observing conditions as $x_{\text{ON}}[n]$ (i.e. similar spatial radio environment), and large enough to avoid any side lobe leakage from a potential ``strong'' extra terrestrial transmitter. If the telescope spatial response is sufficiently known, an appropriate angular offset is the first null between the primary and first secondary lobe.
\end{itemize}

The binary hypotheses problem becomes:

\begin{align*}
&x_{\text{ON}}[n]  = \epsilon \cdot x_{\text{RFI}}[n] + x_{\text{noise}}[n]
{\begin{cases}
{} & \emph{H0}\\
{}+ x_{\text{ET}}[n] & \emph{H1}
\end{cases}}\\
&x_{\text{OFF}}[n] = x_{\text{RFI}}[n] + x_{\text{noise}}[n]
\end{align*}
with $\epsilon \in \mathbb{C}$ corresponding to the RFI gain variation between the \emph{ON}- and \emph{OFF}-target directions.
For short observation cadence (of the order of minutes) and small angle offset between the \emph{ON}- and \emph{OFF}-steering positions, $\epsilon\approx 1$. $\epsilon$ may be inferred when both the RFI environment and the dish telescope radiation pattern are known.

\subsection{$\mathcal{F}$-ratio test for single-dish SETI}
\label{sec:ftest}

The $\mathcal{F}$-ratio test $\theta_{\mathcal{F}}$ is built following:

\begin{equation}
\theta_{\mathcal{F}} = \hat{\sigma}_{\text{ON}}^2 / \hat{\sigma}_{\text{OFF}}^2
\end{equation}
where $\hat{\sigma}_{\text{ON}}^2$ and $\hat{\sigma}_{\text{OFF}}^2$ are evaluated according to equation \ref{eq:estimate} for $x_{\text{ON}}[n]$ and $x_{\text{OFF}}[n]$, respectively.

The normalized $\mathcal{F}$ ratio test
follows a Fisher $\mathcal{F}$ distribution \cite{distributions}. The four data models considered in section \ref{sub:dataproduct} are distributed as:

\begin{itemize}[noitemsep]
\item Wide band RFI / Wide band ET:\\
$\theta_{\mathcal{F}_{\emph{H1}}} \sim \left\{\frac{\sigma_{\text{ET}}^2+\epsilon \cdot \sigma_{\text{RFI}}^2+\sigma_{\text{noise}}^2}{\sigma_{\text{RFI}}^2+\sigma_{\text{noise}}^2}\right\}\mathcal{F}(N,N)$\\
$\theta_{\mathcal{F}_{\emph{H0}}} \sim \left\{\frac{\epsilon \cdot \sigma_{\text{RFI}}^2+\sigma_{\text{noise}}^2}{\sigma_{\text{RFI}}^2+\sigma_{\text{noise}}^2}\right\}\mathcal{F}(N,N)$
\item Wide band RFI / Narrow band ET:\\
$\theta_{\mathcal{F}_{\emph{H1}}} \sim \left\{\frac{\epsilon \cdot \sigma_{\text{RFI}}^2+\sigma_{\text{noise}}^2}{\sigma_{\text{RFI}}^2+\sigma_{\text{noise}}^2}\right\}\mathcal{NCF}(E_{\text{ET}},N,N)$\\
$\theta_{\mathcal{F}_{\emph{H0}}} \sim \left\{\frac{\epsilon \cdot \sigma_{\text{RFI}}^2+\sigma_{\text{noise}}^2}{\sigma_{\text{RFI}}^2+\sigma_{\text{noise}}^2}\right\}\mathcal{F}(N,N)$
\item Narrow band RFI / Wide band ET:\\
$\theta_{\mathcal{F}_{\emph{H1}}} \sim \left\{1+\text{SNR}\right\}\mathcal{DNCF}(\epsilon^2 E_{\text{RFI}},E_{\text{RFI}},N,N)$\\
$\theta_{\mathcal{F}_{\emph{H0}}} \sim \mathcal{DNCF}(\epsilon^2 E_{\text{RFI}},E_{\text{RFI}},N,N)$
\item Narrow band RFI / Narrow band ET:\\
$\theta_{\mathcal{F}_{\emph{H1}}} \sim \mathcal{DNCF}(\epsilon^2 E_{\text{RFI}} + E_{\text{ET}},E_{\text{RFI}},N,N)$\\
$\theta_{\mathcal{F}_{\emph{H0}}} \sim \mathcal{DNCF}(\epsilon^2 E_{\text{RFI}},E_{\text{RFI}},N,N)$
\end{itemize}
where $\left\{\sigma^2\right\}\mathcal{F}(n,n)$ is the normalized $\mathcal{F}$-distribution over $\theta$ with $N$ (double-)degrees of freedom, variable change $\theta \mapsto \theta/\sigma^2$, and appropriate scaling. $\left\{.\right\}\mathcal{NCF}(a,.,.)$ and $\left\{.\right\}\mathcal{DNCF}(a,b,.,.)$ are the Non-Central-$\mathcal{F}$- and Doubly-Non-Central-$\mathcal{F}$-distributions with non-centrality parameters $a$, or $a$ and $b$, respectively. $E_{\text{RFI}}$ is the energy of the RFI signal in the narrow band case. $\text{SNR} = \sigma_{\text{ET}}^2/\sigma_{\text{noise}}^2$ is the Signal-to-Noise Ratio.

\begin{figure}
\centering
\includegraphics[width=.4\textwidth,height=.16\textheight]{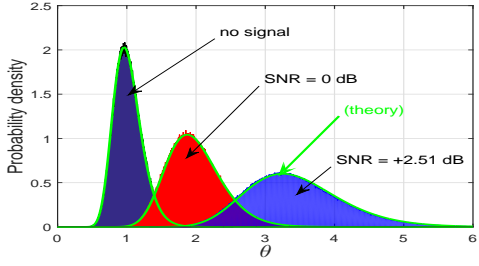}
\caption{$\mathcal{F}$-ratio distributions in the case ``wide band RFI / wide band ET'' with $\epsilon=1$ with real radio telescope data using artificially injected wide band ET transmissions at SNR=0dB and SNR=+2.51dB.}
\label{fig:distribs}
\end{figure}

An example validating the adopted data model is shown in Figure \ref{fig:distribs} where an artificially injected wide band ET transmission has been added to an arbitrary set of \emph{ON} and \emph{OFF} data collected with the Green Bank Telescope (WV, USA)\cite{prestage2009green}. No RFI was found in this data set (i.e. $\sigma_{\text{RFI}}^2=0$).

\subsection{\emph{ON}-\emph{OFF} test for single-dish SETI}
\label{sec:onoff}

The \emph{ON}-\emph{OFF} detector $\theta_{\emph{ON}-\emph{OFF}}$ is constructed as follows:

\begin{equation}
\theta_{\emph{ON}-\emph{OFF}} = \hat{\sigma}_{\text{ON}}^2 - \hat{\sigma}_{\text{OFF}}^2
\end{equation}

$\hat{\sigma}_{\text{ON}}^2$ under \emph{H1} and the various data models considered is distributed as:

\begin{itemize}[noitemsep]
\item Wide band RFI / Wide band ET:\\
$\hat{\sigma}^2_{\text{ON}_{\text{\emph{H1}}}} \sim \chi^2(N,\sigma_{\text{ET}}^2+\epsilon^2  \sigma_r^2+\sigma_n^2)$\\
$\hat{\sigma}^2_{\text{ON}_{\text{\emph{H0}}}} \sim \chi^2(N,\epsilon^2 \sigma_r^2+\sigma_n^2)$
\item Wide band RFI / Narrow band ET:\\
$\hat{\sigma}^2_{\text{ON}_{\text{\emph{H1}}}} \sim \chi^2_{\text{NC}}(N,\epsilon^2 \sigma_r^2+\sigma_n^2,E_{\text{ET}})$\\
$\hat{\sigma}^2_{\text{ON}_{\text{\emph{H0}}}} \sim \chi^2(N,\epsilon^2 \sigma_r^2+\sigma_n^2)$
\item Narrow band RFI / Wide band ET:\\
$\hat{\sigma}^2_{\text{ON}_{\text{\emph{H1}}}} \sim \chi^2_{\text{NC}}(N,\sigma_{\text{ET}}^2+\sigma_n^2,\epsilon^2 E_{\text{RFI}})$\\
$\hat{\sigma}^2_{\text{ON}_{\text{\emph{H0}}}} \sim \chi^2_{\text{NC}}(N,\sigma_n^2,\epsilon^2 E_{\text{RFI}})$
\item Narrow band RFI / Narrow band ET:\\
$\hat{\sigma}^2_{\text{ON}_{\text{\emph{H1}}}} \sim \chi^2_{\text{NC}}(N,\sigma_n^2,E_{\text{ET}} + \epsilon^2 E_{\text{RFI}})$\\
$\hat{\sigma}^2_{\text{ON}_{\text{\emph{H0}}}} \sim \chi^2_{\text{NC}}(N,\sigma_n^2,\epsilon^2 E_{\text{RFI}})$
\end{itemize}
where $\chi^2(N,\sigma^2)$ is the central $\chi^2$ distribution with $N$ degrees of freedom and power $\sigma^2$, and $\chi^2_{\text{NC}}(N,\sigma^2,a)$ is the non-central $\chi^2$ distribution with $N$ degrees of freedom, power $\sigma^2$ and non-centrality parameter $a$. $\hat{\sigma}^2_{\text{OFF}_{\text{\emph{H1}}}}$ and $\hat{\sigma}^2_{\text{OFF}_{\text{\emph{H0}}}}$ are distributed as $\hat{\sigma}^2_{\text{ON}_{\text{\emph{H0}}}}$ with $\epsilon^2 = 1$.

The probability distribution of $\theta_{\emph{ON}-\emph{OFF}}$ under \emph{H0} and \emph{H1} is a difference of independent $\chi^2$ distributions, as described in \cite{prob_distrib_gaussian_variables}.

\subsection{Detectors comparison}

The detection performance of a given detector $\theta$ depends on the statistical distribution of the test under both hypotheses \emph{H0} and \emph{H1}.

For a given data model and threshold $\tilde{\theta}$, the detection and false alarm probabilities, $\mathcal{P}_d$ and $\mathcal{P}_{fa}$, are given by:

\begin{equation}
\mathcal{P}_d = \int_{\tilde{\theta}}^{\infty} \theta_{\emph{H1}} d\theta \ \ \ \ \ \ \ \ \mathcal{P}_{fa} = \int_{\tilde{\theta}}^{\infty} \theta_{\emph{H0}} d\theta
\end{equation}

\begin{figure}[h]
\centering
\subfigure[]{
\includegraphics[height=0.13\textheight]{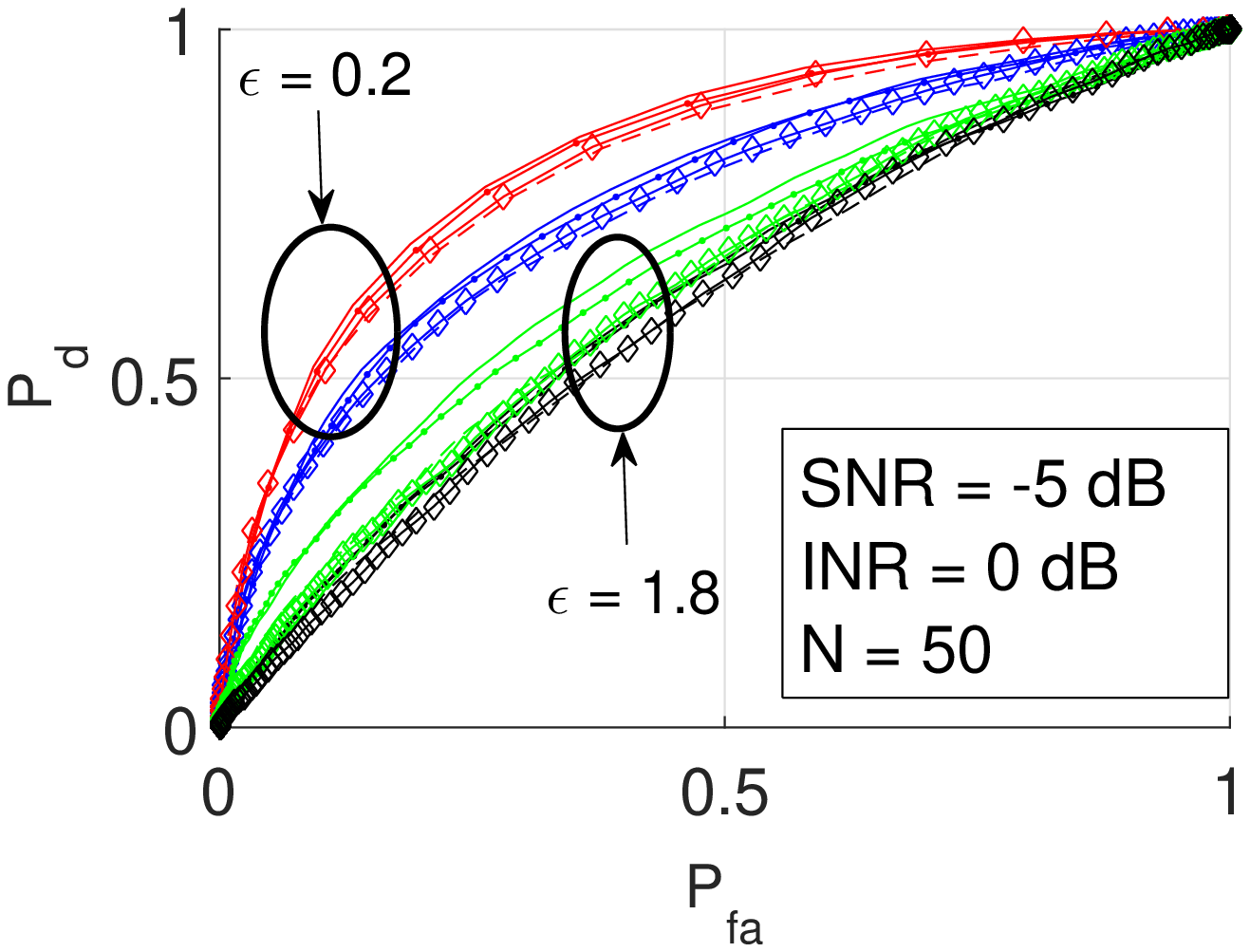}
}
\subfigure[]{
\includegraphics[height=0.13\textheight]{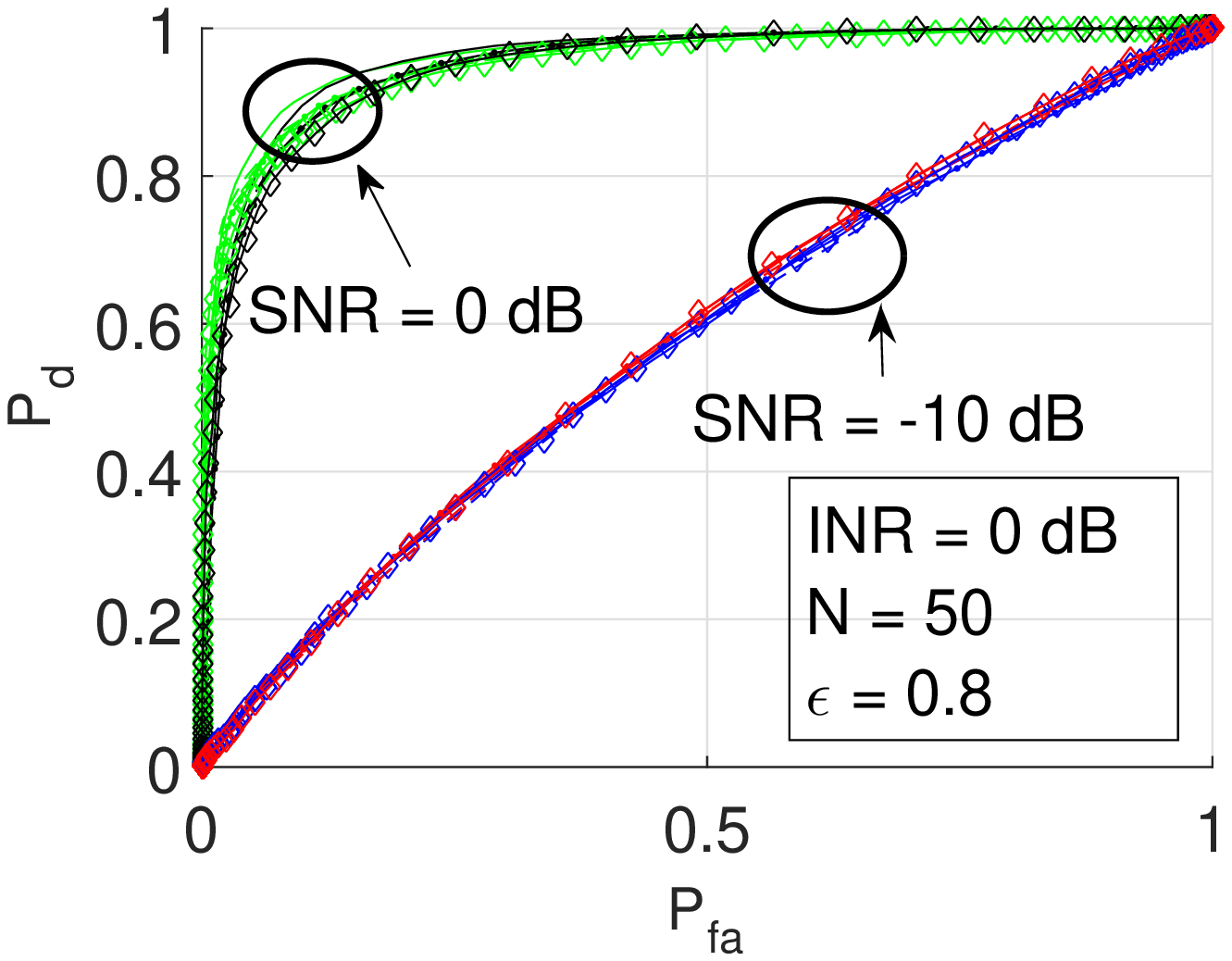}
}
\caption{Receiver Operating Characteristics. Probability of detection ($\text{P}_\text{d}$) vs. probability of false alarm ($\text{P}_\text{fa}$). Red and green lines : $\theta_{\emph{ON}-\emph{OFF}}$. Black and blue lines : $\theta_{\mathcal{F}}$. (\textit{plain line}) narrow band RFI / narrow band ET. (\textit{dashed line}) wide band RFI / narrow band ET. (\textit{dotted line}) narrow band RFI / wide band ET. (\textit{diamond}) wide band RFI / wide band ET.}
\label{fig:roc}
\end{figure}

Those quantities enable the comparison of the detectors under variable data model parametrization through the evaluation of Receiver Operating Characteristic (ROC) curves \cite{van2004detection}, as shown on Figure \ref{fig:roc}.

The following observations follow the ROC analysis of both detectors:

\begin{itemize}[noitemsep]
\item The data model (narrow / wide band) does not significantly affect the detectors performance.
\item The $\emph{ON}-\emph{OFF}$ approach is less impacted by $\epsilon$ variations than the $\mathcal{F}$-ratio approach (Figure \ref{fig:roc}.(a)).
\item Low SNR significantly impact the detection performance, and signals with SNR below -10 dB are undetectable using the power approaches.
\end{itemize}

\section{Conclusion}
\label{sec:conclusion}

Ground based single dish telescopes offer appropriate sensitivity and computational resources to conduct SETI observations, but standard astronomical data product are limited to energy estimates.
This paper formulates the statistical data model of a SETI experiment with such an instrument, and presents two detection strategies overcoming the classical energy detection caveats.
The related test statistics are derived to enable a theoretical detection performance assessment.


The proposed approaches require asynchronous \emph{ON} and \emph{OFF}
target observations. While spatial and temporal stationarity are assumed, spillover noise and time-dependent uncalibration can affect the performance of the detectors, as well as RFI transmission paths variations between the steered directions.
Experiments involving $\emph{ON} / \emph{OFF}$ observations of artificial extra terrestrial sources, such as the Voyager 1 spacecraft, will be necessary to validate the proposed approaches and infer their sensitivities.

A natural extension of the proposed detection schemes consists in multi hypothesis testing, addressing the case of an additional extra-terrestrial transmission present in the $\emph{OFF}$ measurement.

\bibliographystyle{IEEEbib}
\bibliography{refs}

\end{document}